\documentclass[a4paper]{jpconf}
\usepackage{graphicx}
\graphicspath{{images/}}
\usepackage{amssymb, amsmath}
\usepackage{hyperref}
\begin{document}
\title{Equation of state and composition of the inner crust of an accreting neutron star: multicomponent model}

\author{N N Shchechilin$^{1,2}$ and A I Chugunov$^2$}

\address{$^1$~Peter the Great St.\ Petersburg Polytechnic University, Politekhnicheskaya 29, 195251 St. Petersburg, Russia}
\address{$^2$~
 Ioffe Institute, Politekhnicheskaya 26, 194021 St. Petersburg, Russia}

\ead{nicknicklas@mail.ru}

\begin{abstract}
Correct interpretation of X-ray observations of transiently accreting neutron stars requires modeling of nuclear-physical processes in these objects. We consider a chain of nuclear reactions that drives the crust composition in an accreting neutron star and heats up the star. We constructed multicomponent approach with the kinetics of nuclear reactions described in simplified stepwise manner. The redistribution of nucleons between nuclei by emission and capture of neutrons is shown to significantly affect the nuclear reaction chains and the composition of the inner crust. In particular, even if the outer crust has one-component composition, the appearance of free neutrons in the inner crust leads to branching of reaction chains and formation of the multicomponent composition. We apply the compressible liquid drop nuclear model, which includes effects of free neutrons on nuclear energies. It allows us to calculate the composition, the heating profile and the equation of state of matter up to densities $\rho\simeq 2\times 10^{13}$~g\,cm$^{-3}$.       
\end{abstract}

\section{Introduction}
Modern telescopes observe neutron stars in whole range of an electromagnetic emission (from radio to gamma rays), furthermore the recent detection of gravitational waves from the neutron star merger opens era of multimessenger observations of neutron stars \cite{GW170817}. The observational data are applied to test numerous theoretical models of the superdense matter in the depths of a neutron stars (e.g.\  \cite{Baym19}). Especially interesting objects are neutron stars in binary systems, in particular in Low Mass X-ray Binaries (LMXBs), where companion of the neutron star is a Roche-lobe-filling low mass star ($\approx M_\odot$). In LMXBs a matter is transfered from the companion star to the neutron star via an accretion disk. This process leads to the strong X-ray luminosity, which is well detected by X-ray telescopes.

Arriving to the neutron star surface, the matter is buried 
by the new-coming mass from the disk and eventually replaces the pristine neutron star crust. During this process nuclei in the accreted matter undergo nuclear reactions and the crust of the neutron star is heated up \cite{BBR}. For several sources with transient (non-stationary) accretion, 
the thermal X-ray emission of the neutron star surface was detected during quiescent periods \cite{WIJNANDSetal17}. It allows to analyze the thermal relaxation of the crust-core system, 
estimate the temperature of the core and put constraints on a crust and core composition, processes of neutrino emission, 
neutron superfluidity, etc. \cite{MeiselShterninetal18}.

The neutron star crust acts a significant role in the thermal evolution of LMXBs and should be modeled precisely to arrive to the correct interpretation of the observational data.
In particular, detailed calculations of the nuclear reaction sequence, which drives the crust composition, are required, because 
due to the relatively low temperatures
($T\lesssim 5\times10^8$~K)
the nuclear fusion is suppressed, and the matter does not arrive to the thermodynamic equilibrium with respect to the nuclear composition. 
First calculations were made by Sato \cite{Sato79}, the most widely used model was suggested by  Haensel \& Zdunik \cite{HZ90,HZ08}. In the recent work Lau et al.\ \cite{Lauetall18} constructed the reaction network model for the neutron star crust up to densities $2\times10^{12}$~g\,cm$^{-3}$. Here we present the model applicable up to $\rho\lesssim 2\times 10^{13}$~g\,cm$^{-3}$.

\section{Multicomponent model}
Our model in general follows ref.\ \cite{HZ90,HZ08}, hence it considers nuclear reactions in the accreted matter on course of an increasing pressure. However, we discard the simplifying assumption about the one-component composition of the matter at fixed pressure and take into account a possibility of neutron exchange between atomic nuclei. As in ref.\ \cite{HZ90}, a nuclear binding energy is calculated within the Mackie \& Baym compressible liquid drop model \cite{MB77}, which includes effects of free neutrons on the nuclei surface tension.
Following \cite{HZ90,Steiner2012}, we neglect finite temperature effects ($T=0$) and set the initial composition to be pure ${\rm{^{56}Fe}}$,
which supposed to be the dominating element in a superbursts ash \cite{MeiselShterninetal18}.

Following \cite{Steiner2012}, we apply the simplified model for the kinetics of nuclear reactions:
at each step only a ``chunk'' ($\alpha = 0.01$ of total nuclei amount) of nuclei undergoes energetically allowed nuclear reaction.
Reactions are ``offered''
in the following order:%
\footnote{Ref.\ \cite{Steiner2012} does not contain a detailed description of the priority rules for allowed reactions, thus a detailed comparison of the kinetic model is not possible. 
\label{Steiner_model}
}
(a) neutron emission, (b) 2 neutrons emission, (c) neutron capture, (d) 2 neutrons capture, (e) electron capture, (f) pycnonuclear fusion. If some reaction decreases Gibbs energy, it proceeds for the fraction $\alpha$, and the program returns to the reaction (a). 
The order of ``offering'' correspond to general order-of-magnitude estimates of typical reaction timescales: the nuclear timescale ($\sim 10^{-21}$~s)  for emission of unbound neutron(s);  picoseconds for exothermic neutron captures \cite{Gorielyetal08};  of order of $10^{-5}-10^3$~s for electron captures, and much longer for pycnonuclear reactions.
The preference between rates of (d) and (e) reactions is not so certain: within our approach the  reaction of type (d) can take place, only if the one neutron capture 
is not energetically favorable and require a thermal excitation.
However, if the subsequent neutron capture is energetically favorable, it takes place very fast, leading to 2 neutron capture as a net result. To take into account this possibility we include reactions of type (d) to our simplified kinetic approach. Its rate is primarily determined by the rate of the first neutron capture, which should  be calculated including plasma and neutron degeneracy effects \cite{Shterninetal12}. The resulting rate strongly depends on a neutron capture threshold and a neutron chemical potential. Here it is assumed to be faster than the electron capture rate for simplicity.
Finally, if reactions of the same type are allowed for several types of nuclei, we assume that the reaction  (a-d) takes place for the most abundant element; while
among allowed electron capture reactions we select the most energy efficient one to proceed first.  
The pycnonuclear fusion is switched on for elements with $Z$ equal to 10 and 12 at the pressure thresholds, which corresponds to pycnonyclear reaction for this $Z$ as given in \cite{HZ08}.
If all considered reactions can not decrease Gibbs energy, 
we increase the pressure in a stepwise manner.
If after the next step at least one reaction becomes allowed, we adjust pressure accurately to the threshold 
of the first allowed reaction and apply the procedure described above for the adjusted pressure.
By construction, our simplified kinetics model is independent of an accretion rate.

We assume quasineutrality ($n_e=\sum_i Z_i n_{N_i}$) and a uniform distribution of free neutrons [$Y_n n_b={n_n}\left(1-\sum_i V_{N_i}X_in_N\right)$].
Here  $n_n$, $n_e$, $n_N$, $n_b$ are number densities of neutrons, electrons, nuclei and baryons respectively;
$Z_i, V_{N_i}, X_i$ being charge, volume and abundance of certain type of nuclei, $Y_n$ denotes the free neutrons abundance.
The total pressure calculated as $P=P_e+P_l+P_n$,
where $P_e, P_n, P_l$ denote pressures of electrons, free neutrons and lattice.

\section{Results}
 
\subsection{Crust composition}

\begin{figure}[t]
	\includegraphics[width=38pc]{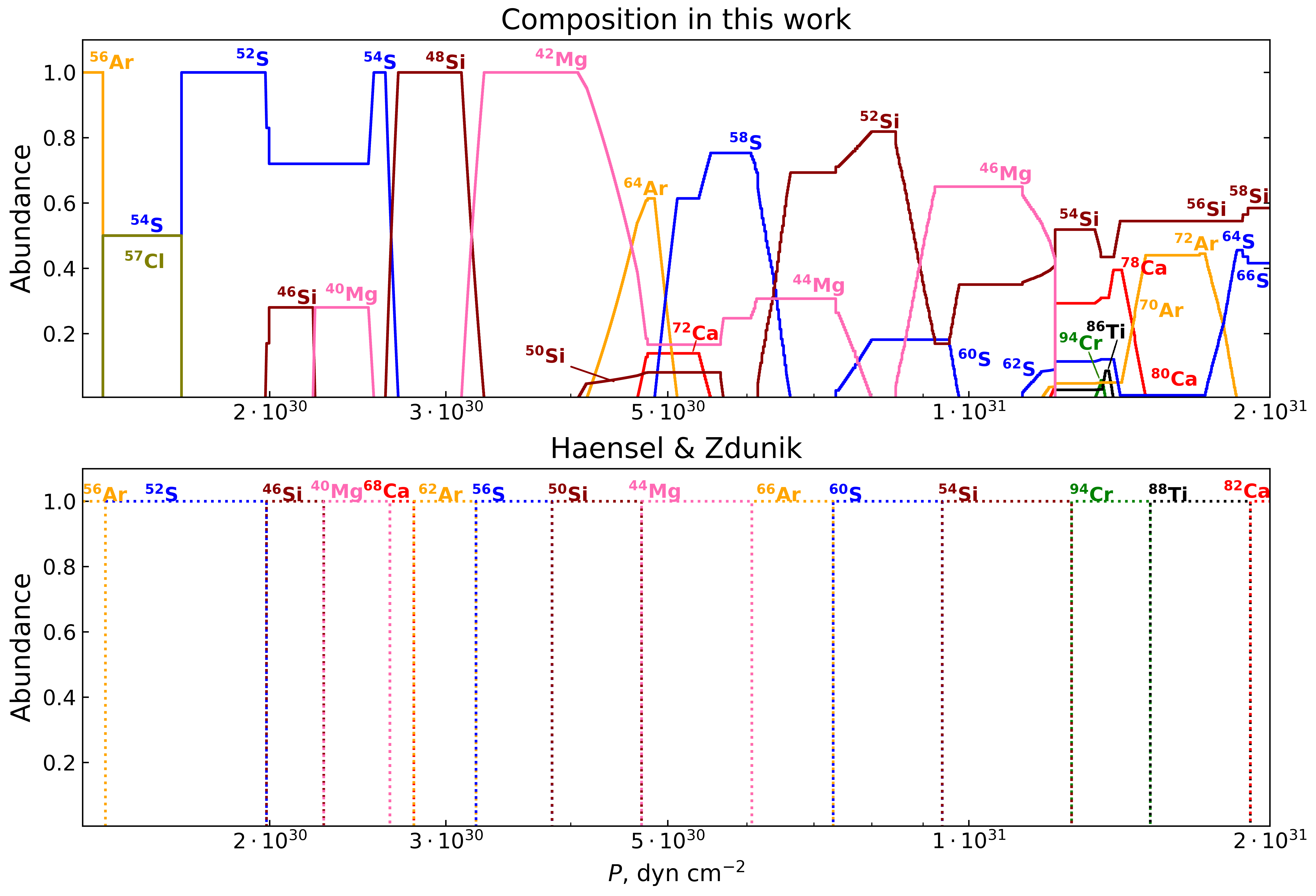}
	\caption{\label{Evol1} The composition of the accreted crust in our model (upper panel) and from ref.\ \cite{HZ08} (lower panel). The shown abundances are the sum over all isotopes; the most abundant isotope is indicated.}
\end{figure}

Figure \ref{Evol1} demonstrates the composition of the accreted inner crust. By definition free neutrons are absent in the outer crust and thus only reactions of type (e) are allowed there. As as result, our model predicts the same composition as ref.\ \cite{HZ08}. In particular, the neutron drip point is $\rho=6.11\times 10^{11}$~g\,cm$^{-3}$. In contrast to \cite{HZ08}, at the neutron drip the fraction of $\rm{^{56}Ar}$ undergoes
$(n,\gamma)$ reaction, followed by the electron capture, leading to formation of ${^{57}\rm{Cl}}$, which is $\beta$-stable at given pressure\footnote{The Mackie \& Baym model appeals to the energy density of pure neutron matter 
from \cite{Sjoberg74}, which overestimate the energy at low number density of neutrons. It leads to an unphysical suppression of neutron emission by a small fraction of nuclei. To avoid this feature we set $\alpha=0.5$ at the neutron drip.}.
The subsequent transitions from one type of nuclei to another are accompanied by gradual increase of the pressure (in \cite{HZ90,HZ08} transitions take place at the fixed pressure and are accompanied by the density jump). As in \cite{Lauetall18}, heavy nuclei formed during pycnonuclear fusion are unstable and pass through neutron emissions and electron captures. This results in an effective conversion of nuclei to neutrons and a substantial energy release. Pycnonuclear reactions proceed in certain pressure intervals, whereas in \cite{HZ90,HZ08} pycnonuclear fusions occur at the fixed pressure for all nuclei. The average charge of nuclei in our multicomponent model turns out to be less than in \cite{HZ08}.
In our model the impurity parameter is rather small (never exceeds $\lesssim10$, having pronounced peaks in regions where pycnonuclear reactions are active).
It qualitatively agrees with results by \cite{Lauetall18} at a respective density region, but in a bit of a contrast with results of Ref.\ \cite{Steiner2012}, which predicts the vanishingly small impurity parameter for $\rho< 10^{13}$~g\,cm$^{-3}$, but at larger densities it rapidly increases up to $\sim 20-30$.

\subsection{Equation of state and heating}

\begin{figure}[t]
	\begin{minipage}[t]{18pc}
		\includegraphics[width=18pc]{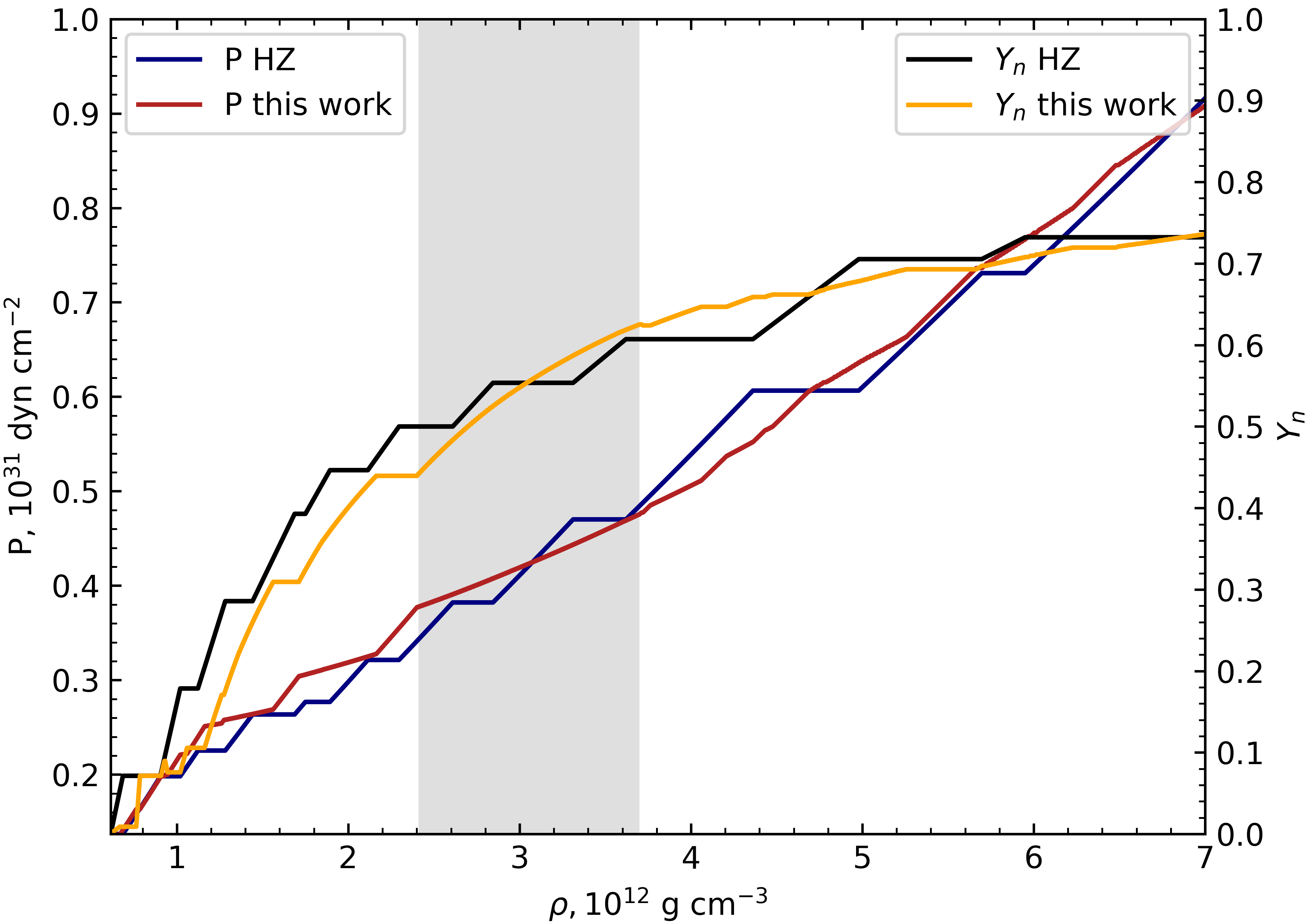}
		\caption{Equation of state and free neutron abundance in the multicomponent model and in the model of ref.\ \cite{HZ08} (HZ). The gray area indicates one of the pycnonuclear reaction intervals in multicomponent model.}
		\label{EoS}
	\end{minipage}
	\hspace{2pc}
	\begin{minipage}[t]{18pc}
		\includegraphics[width=18pc]{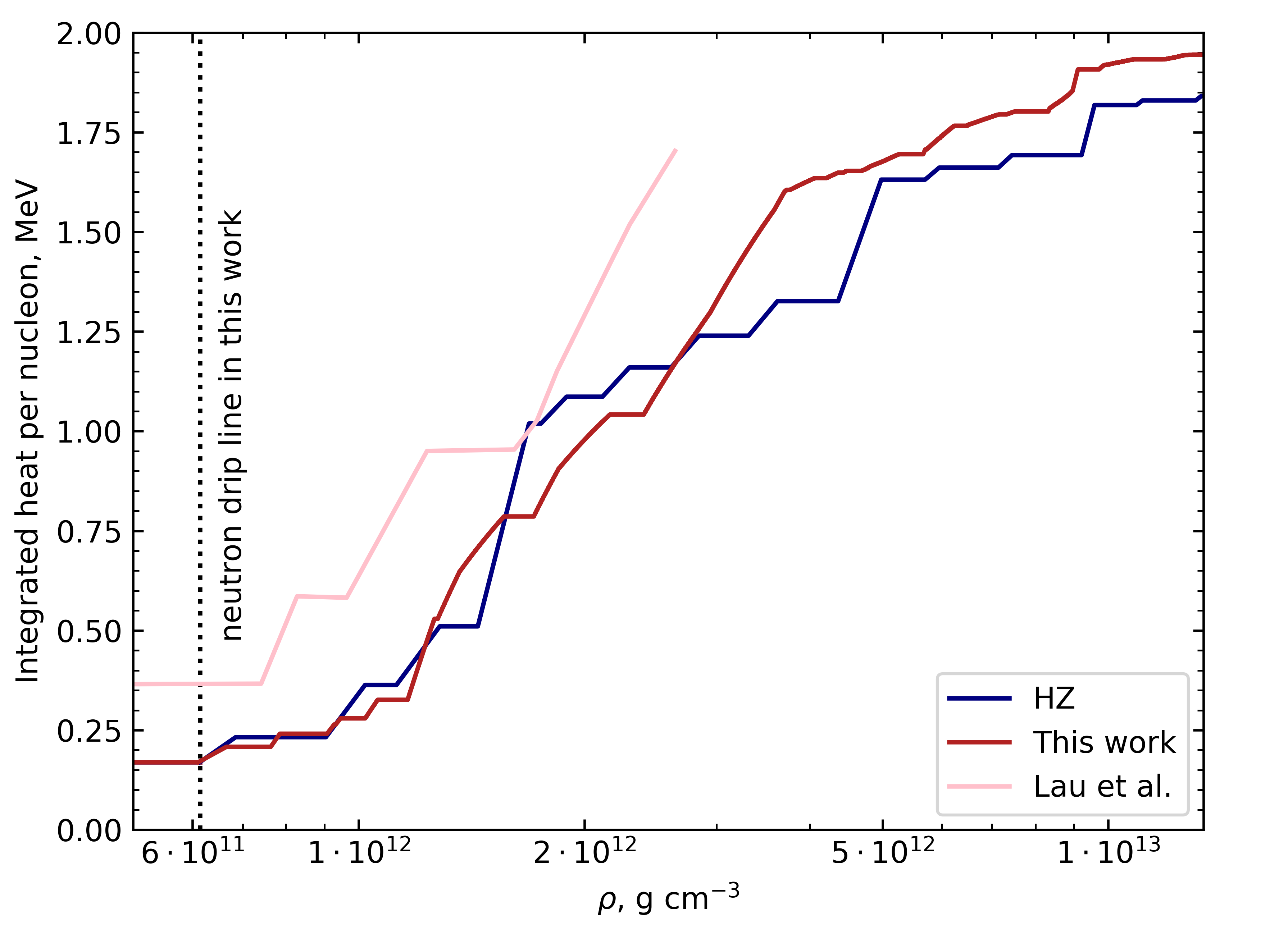}
		\caption{Integrated heat in comparison with the model \cite{HZ08} (HZ) and \cite{Lauetall18} (Lau et al.).
		The vertical dotted line indicates the boundary of the inner and outer crust (the neutron drip point) in this work.}\label{Q} 
	\end{minipage} 
\end{figure}

Equation of state, shown in figure \ref{EoS}, displays smoother behavior and softening in pycnonuclear intervals in comparison with \cite{HZ08}. It correlates with the appearance of larger amount of free neutrons. At densities $7\times 10^{12}$~g\,cm$^{-3}$ the free neutron abundance is about 75\% and neutrons give major contribution to the pressure; the equation of state becomes similar for both models.  

The energy release in nuclear reactions is a crucial parameter for models of the thermal evolution. In weak reactions $(e^-,\nu_e)$ a neutrino carries away a fraction of energy, however, as shown in \cite{Guptaetal07}, transition to an excited state of a daughter nuclei reduces the energy leakage. Following \cite{HZ08}, we neglect neutrino losses. Integrated heat
 (the total energy release in layers with the density below $\rho$) 
is shown in figure \ref{Q}. The multicomponent approach leads to the smooth profile of the energy release and a bit stronger total heating
($\sim 2$~MeV per accreted nucleon at $\rho=2\times10^{13}$~g\,cm$^{-3}$) in comparison with \cite{HZ08}. Ref.\ \cite{Lauetall18} predicts a larger heating power at respective density region, which is likely associated with usage of another nuclear mass model in this reference. 

\section{Summary and conclusions}
We construct the multicomponent model of the inner crust of an accreting neutron star. Even for the same initial composition (pure $^{56}$Fe) results of our model differ from the one-component model of ref.\ \cite{HZ08}. Namely, the composition of the inner crust is predicted to be multicomponent and transitions of one type elements to another in the inner crust occur gradually (without strong density jumps). There are also distinctions in reaction sequences and dominating nuclear elements. Neutron captures play important role in the inner crust and lead to branching of nuclear reaction chains. The smoothed heating profile and the impure composition of the inner crust can affect models of thermal evolution, which are crucial for the interpretation of the observational data on transiently accreting neutron stars. The smoothness of the equation of state is important for a neutron star seismology.
Even at the highest density of our model $\rho=2\times10^{13}$~g\,cm$^{-3}$ the nuclear composition does not reach the ground state.

The multicomponent composition and neutron captures also were discussed in ref.\ \cite{Steiner2012}, as well in ref. \cite{Guptaetal08} and a subsequent paper \cite{Lauetall18}. The reaction network model of \cite{Lauetall18} includes the detailed kinetics of nuclear reactions (with realistic reaction rates), but restricted to the density $\rho\lesssim 2\times 10^{12}$~g\,cm$^{-3}$. Here we construct the model applicable up to $2\times 10^{13}$~g\,cm$^{-3}$. We use the stepwise approach similar to \cite{Steiner2012}. It captures main features of the detailed reaction network in a simplified, but physically clear basis. In particular, it describes the role of neutron captures in producing multicomponent mixtures and the conversion of nuclei into free neutrons, triggered by pycnonuclear reactions. In contrast to \cite{Steiner2012} we employ the Mackie \& Baym nuclear model \cite{MB77} and provide detailed description of our kinetic model. It is worth to stress, that our approach can be easily generalized for an arbitrary mass model.

\ack
We are grateful to D G Yakovlev, M E Gusakov,  K P Levenfish, and P S Shternin for useful discussions. Work is supported by Russian Science Foundation (grant 19-12-00133).

\end{document}